# Identifying Selections Operating on HIV-1 Reverse Transcriptase via Uniform Manifold Approximation and Projection

Code available at github.com/shefaliqamar/HIV-Epistatic

SHEFALI QAMAR

University of California, Santa Cruz; sqamar@ucsc.edu

MANEL CAMPS

University of California, Santa Cruz; mcamps@ucsc.edu

JAY KIM

University of California, Santa Cruz; jay.wj.kim@gmail.com

We analyze 14,651 HIV1 reverse transcriptase (HIV RT) sequences from the Stanford HIV Drug Resistance Database labeled with treatment regimen in order to study the evolution this enzyme under drug selection in the clinic. Our goal is to identify distinct sectors of HIV RT's sequence space that are undergoing evolution as a way to identify individual selections and/or evolutionary solutions. We utilize Uniform Manifold Approximation and Projection (UMAP), a graph-based dimensionality reduction technique uniquely suited for the detection of non-linear dependencies and visualize the results using an unsupervised clustering algorithm based on density analysis. Our analysis produced 21 distinct clusters of sequences. Supporting the biological significance of these clusters, they tend to represent phylogenetically related sequences with strong correspondence to distinct treatment regimens. Thus, this method for visualization of areas of HIV RT undergoing evolution can help infer information about selective pressures, although it is correlative. The mutation signatures associated with each cluster may represent the higher-order epistatic context facilitating these evolutionary pathways, information that is generally not accessible by other types of mutational co-dependence analyses.

# 1 INTRODUCTION

The Human Immunodeficiency Virus (HIV) is a retrovirus that is the causative agent of Acquired Immunodeficiency Syndrome (AIDS), currently infecting over 36 million adults and 1.7 million children. Advances in antiretroviral therapy allow patients with HIV to live long lives while hosting the virus. However, thanks to its extremely high mutation rate, HIV-1 has evolved resistance to some of these antiretroviral drugs and in 2020 AIDS was still causing 680,000 deaths worldwide [1].

HIV-1 infects primarily CD4+ T cells by injecting viral RNA along with reverse transcriptase, protease, and integrase proteins. The single-stranded viral RNA is transcribed into double-stranded DNA by the reverse transcriptase (RT) and integrated into the cell's genome by the integrase. The functional expression of viral genes requires processing by a viral protease [2]. All these processes and the fusion of the viral envelope with the host's cell membrane can be inhibited by drugs. Here, we focus primarily on HIV RT inhibitors, which represent a major class of antiretrovirals [20]. These come in two flavors. The first class of antiretroviral drugs are nucleoside reverse transcriptase inhibitors (NRTIs), which are nucleoside analogs that compete with dNTPs for incorporation, inhibiting reverse transcriptase. The second class are non-nucleoside reverse transcriptase inhibitors (NNRTIs), which bind at the hydrophobic pocket, interfering with DNA synthesis [4,5].

RT replication is error prone and represents a major source of mutations [19]. Given that most random mutations are deleterious, replicating HIV genomes are under constant purifying selection [3]. Under positive selective pressure by HIV-RT inhibitors, drug resistance mutations (DRMs) are selected. Some of these DRMs have been identified as diagnostic for resistance to NRTIs and to NNRTIs and are routinely used for predictive diagnostics [16]. Well characterized NRTIs resistance mutations are found at positions 41, 65, 67, 70, 74, 115, 184, 210, 215, and 219 [4], and diagnostic NNRTIs resistant mutations are found at positions 100, 101, 103, 106, 138, 181, 188, 190, and 230 [5].

NRTIs and NNRTI are often prescribed in combination [6, 17]. More specifically, highly-active antiretroviral therapy (HAART) is a drug cocktail treatment used to reduce HIV viral load sufficiently to prevent the progression to AIDS. HAART includes at least three medications, namely one protease inhibitor (PI) and two to three reverse transcriptase inhibitors. A selection with drugs acting by different mechanisms minimizes the probability of resistance and when resistance evolves maximizes the probability of having a high fitness cost. Decreased fitness associated with adaptive DRMs leads to the selection of compensatory mutations, generating networks of co-dependent mutations. The landscape of the sequence context supporting the evolution of gain-of-function mutations has not been studied in all its constitutive interdependencies, particularly beyond pairwise interactions.

Dimensionality reduction methods such as multi-dimensional scaling (MDS) and principal component analysis (PCA) have been previously reported [7]. These analyses define a pairwise distance metric along positions on the HIV genome in order to create a distance matrix, then perform an eigenvector decomposition on this matrix to observe how the mutations relate in lower-dimensional space. However, a linear dimensionality reduction limits the identification of large, interdependent networks of mutations that may be evolving together. Indeed, a more recent study performing an integrated analysis of residue coevolution and protein structures showed complex interdependencies between mutations [21]. Here we turn to a non-linear dimensionality reduction technique, which allows for identification of relationships between an unbounded number of mutations simultaneously, with the hope of capturing such interdependencies.

The approach that we used is Uniform Manifold Approximation and Projection (UMAP) [8]. UMAP is a graph-based dimensionality reduction technique that is uniquely suited for the detection of non-linear dependencies. Amongst graph-based algorithms, UMAP is known to have superior preservation of local structure because it implements a gravity-like pull for closeby points. This is advantageous to the analysis of non-linear dependencies between mutations, where the number of co-dependent mutations that can be analyzed is unbounded. UMAP enables sequences with similar nonlinear dependencies between mutations to be drawn closer together because it prioritizes these local interactions. This drew a compelling case for applying UMAP directly to the RT sequence data.

Our curated database consisted of 14,651 HIV1 reverse transcriptase sequences from the Stanford HIV Drug Resistance Database labeled with the treatment regimen that patients were undergoing at the time of viral isolation. Our analysis produced clusters of sequences showing distinct sectors of HIV RT's sequence space that are accessible to evolution. These clusters tended to be phylogenetically related, and we found a strong correspondence between how homogenous the sequence was within a sector and how consistent the treatment that the patient had received was. This suggests that this method can be used to infer information about the number and nature of selections driving the evolution of a protein. Strikingly, antiretroviral cocktails but the known adaptive DRMs are largely absent in the mutation profiles associated with each of these clusters. We suggest that the mutations consistently found in these clusters represent distinct configurations of higher- order epistatic interactions that determine the accessibility of RT sequence space to evolution.

## 2 METHODS

### 2.1 Dataset

The 72,200 HIV-1 reverse transcriptase sequences used in this analysis were sourced from the Stanford HIV Drug Resistance Database [9]. Primarily, these sequences mapped the p66 subunit of HIV-1 reverse transcriptase, consisting of 560 amino acids and its enzymatic activities. These sequences were taken from patients involved in longitudinal studies from patients under different treatment regimens. For each sequence, the following information was also given: patient ID, study ID, study year, and a list of the drugs administered to the patient for the purpose of HIV treatment prior to sequencing. Here, we will refer to this additional information frequently as the sequence's metadata.

The number of sequences per patient varied depending on both the study and how many drugs were introduced into the treatment regimen over time. Further, the various studies represented chose to sequence different regions and amounts of the reverse transcriptase, which raised the need for multiple sequence alignment.

### 2.2 Sequence Alignment and Reference Sequence

Prior to alignment, the sequences were translated from nucleotide sequences into amino acid sequences via the Biopython package's Sequence module [10]. During translation, 34,177 sequences which included stop codons were identified and filtered out of the dataset to be aligned, as these represented abortive viruses. A further 510 sequences were filtered out due to errors in translation. The remaining 41,529 sequences were compiled into one FASTA file to allow for sequence alignment. These remaining sequences represented 35,873 patients in total.

Multiple sequence alignment was performed on the resulting amino acid sequences using MAFFT: a multiple alignment program for amino acid or nucleotide sequences [11]. The resulting alignment consisted of 563 amino acids, starting at and extending from the first amino acid of the p66 subunit.

The reference sequence used in this analysis was also sourced from the Stanford HIV Drug Resistance Database [9] and was complete for the entire 563 amino acid region spanned by the sequence alignment. This rendered it a complete and suitable reference for the sequences from all of the studies given in the dataset.

### 2.3   Quality Control for Sequences

We designed a method to collapse the database so that each sequence represents the most recent sample from each patient, reasoning that it is the sample that has had the longest time to evolve. Further, the sequences were cropped to focus on the active region of the p66 subunit, consisting of the first 230 amino acids in the alignment. Sequences which were not complete for this region, identified by a dash or 'X' in the alignment, were removed from the dataset. These filters in combination brought the total size of the dataset down to 14,651 aligned, complete sequences, one per patient.

### 2.4   Training Data Encoding and Separation of Metadata

The data was then prepared for analysis by splitting into the training data and metadata. The training data was the only data that was used in the analysis, while the metadata was stored in a separate dataset, to be used after the analysis to interpret results. Separate dataframes were used to store the information but could be cross-indexed in order to obtain information for a given sequence.

Training data consisted of only the sequences. In order to perform numerical analyses on this type of data, a one-hot encoding was utilized to create a matrix representation of the raw sequence data. In this representation, a column was included for each of the possible amino acids present at each position along the sequence (i.e. columns "0_A", "0_F", "0_H", "0_L", "0_P", ….).

Metadata consisted of the patient ID, study ID, study year, raw sequence, and a list of the drugs administered to the patient for the purpose of HIV treatment. This was expanded to include a list of the mutations identified in each sequence. Depending on the sequence of mutations found, sequences were annotated into drug categories to help interpret the correspondence between the communities identified by our unsupervised clustering algorithm based on density analysis and likely selections driving them (see *Density-Based Spatial Clustering of Applications with Noise* below). If a majority of these mutations belonged to either known nucleoside reverse transcriptase inhibitor (NRTI) diagnostic mutations or to known non-nucleoside reverse transcriptase inhibitor (NNRTI) diagnostic mutations, they were labeled "TAM" or "NNRTI", respectively. If no clear majority of either type was found, we labeled the sequence as "None".

### 2.5   Dimensionality Reduction via Uniform Manifold Approximation and Projection

The novel nonlinear dimensionality reduction algorithm Uniform Manifold Approximation and Projection (UMAP) was utilized to obtain a projection of the training dataset onto lower-dimensional space. The implementation of UMAP used here is given by the algorithm's authors and to perform unsupervised learning of a lower-dimensional representation of the data. The hyperparameter of the number of neighbors required to define a point as an interior point was set to 5, and the random state was fixed to 42 as per computing convention. Note that reasonably adjusting the hyperparameter of the number of neighbors required should yield similar dimensionality reduction.

In this lower-dimensional space, each sequence is represented by a single point. Also note that the Euclidean distance between points in this space cannot represent evolutionary distance as the distances are nonlinearly determined by the mutations in the sequences, where one important mutation or combination of mutations can cause a large shift.

### 2.6 Density-Based Spatial Clustering of Applications with Noise (DBSCAN)

Density-Based Spatial Clustering of Applications with Noise (DBSCAN) [12] was chosen as the clustering algorithm for this analysis as it does not require specification of a fixed number of points (sequences) per cluster. The implementation of DBSCAN utilized here is taken from Sci-kit Learn's cluster package. The minimum number of samples (sequences) for each cluster was set to 30 samples and the epsilon value set to 0.5, values conferred upon by methodical adjustments and observation of the clustering quality. Note that reasonably adjusting these hyperparameters should yield similar clustering. In our analysis, this configuration resulted in 21 distinct clusters of which two were considerably larger than the others. Recall that each point in this space represents a sequence. Thus, each cluster of points represents a group of sequences.

For each cluster, we now had a group of sequences which had been brought together based on some latent factors identified by the algorithms. As a control, we checked if the sequences themselves were the sole basis of the clustering as opposed to some variation between studies. Thus, we verified that the sequences in each cluster were from various studies and patients. For instance, the 33 sequences in Cluster 5 came from 33 different patients and 7 different studies.

Cluster 0 from the initial DBSCAN clustering analysis was very large, consisting of about 10,000 diverse sequences. Thus, it was necessary to further break down this cluster via a repeat analysis in order to extract meaningful sub-clusters. First, we collected the training data corresponding to the sequences in Cluster 0 into a separate data frame, then repeated the same UMAP and DBSCAN analysis on this dataset [8, 12]. This resulted in several sub-clusters, of which one was again exceptionally large but two were distinctly identifiable. Later we will see that the information gained from this increased resolution has important meaning.

### 2.7 Histograms and Homogeneity

For each cluster of patient sequences, two frequency histograms were created: one for the frequencies of the mutations, and one for the frequencies of the drug treatments prescribed to the patients. For convenient comparison, these two histograms were included side-by-side for each cluster in a figure.

We defined here a homogeneity index $\mu_{50} / \mu$ to quantify the consistency with which mutations appear within a cluster's sequences: dividing $\mu_{50}$ the number of mutations with frequency greater than 50% by the total number $\mu$ of mutations observed. We also define a corresponding homogeneity index $\lambda_{50} / \lambda$ to quantify the consistency with which drug treatments appear within a cluster's sequences. Here, $\lambda_{50}$ represents the number of mutations with frequency greater than 50% and $\lambda$ represents the total number of mutations observed.

## 2.8 Computing Phylogenetic Distance with UPGMA

To estimate the level of phylogenetic relatedness of the sequences that were clustered together, we utilized the UPGMA algorithm for building phylogenetic trees from given sequences [14]. The implementation used for UPGMA was from BioPython's Phylo package [10].

First, clusters of random sequences from the dataset were generated in varying sizes as a control. For each of these clusters, a UPGMA tree was built and the tree depth was recorded, which is defined as the maximum distance from the original ancestor to a descendant. Then, for each of our clusters from the DBSCAN clustering, this analysis was repeated and the tree depth was recorded.

## 2.9 Longitudinal Analysis

For each patient, we build a network at the first and last sequencing point. Mutations were called by aligning the last and the first sequence.

## 3 RESULTS AND DISCUSSION

### 3.1 Goal and Approach

The goal of this study is to map out distinct sectors of HIV RT's sequence that are accessible to evolution as a way to get information about types of selections that HIV RT is undergoing in the clinic and about the adaptive solutions found by evolution. Our assumption is that HIV RT's evolution is driven in different directions by a variety of factors (replication efficiency, resistance to drug treatment, fidelity of replication, etc.) while at the same time being shaped by epistasis. We reasoned that the combination of directional selection, which favors the fixation of specific adaptive mutations, and epistasis, which can enhance the effect of adaptive mutations, but which also restricts the accessible sequence space, will allow us to cluster sequences according to either the type of adaptive solution or the type of selection driving them. For this approach to work, we needed two elements: (a) a large number of individual sequences and (b) the ability to maximize our ability to detect functional interactions.

Sequence database: we obtained 72,200 HIV-1 sequences from the Stanford HIV Drug Resistance Database, deposited between 1987 and 2017. These sequences mapped the p66 subunit of HIV-1 RT, which harbors its enzymatic activities [9]. Some of these sequences were part of longitudinal studies, with multiple sequential samples corresponding to individual patients. After putting these sequences through quality control filters and selecting only the last sequence available for each patient the total number of sequences went down to 14,651 (see **Methods**). Thus, our curated sequence database represented the most frequent sequence found in the last sample taken from individual patients.

Maximal ability to detect functional interactions: traditional methods such as multi-dimensional scaling and PCA have been previously used to study RT mutations in the context of antiretroviral selection [7]. These methods are based on eigenvalue decomposition of a distance matrix which is ultimately a linear (otherwise known as pairwise) approach to dimensionality reduction. The expectation was to be able to identify communities corresponding to one or a small number of adaptive mutations for a given adaptive solution and a network of co-dependent mutations. At least it was hoped that mutations conferring resistance to NNRTIs would be found in a separate community relative to mutations conferring resistance to NRTIs, given that the targets and mechanisms of resistance for these two classes of antiretroviral drugs are very different. However, when used on complex sequences recovered from the clinic, these linear methods were only partially successful in segregating communities of NRTI and NNRTI-selected mutations [7].

Recently, more sophisticated non-linear dimensionality reduction approaches have been developed [8, 15]. These methods can capture interactions between multiple mutations simultaneously, thus increasing the number of interactions detected. We hypothesized that by increasing the density of the network of interactions detected, we would be able to improve the resolution of the generated mutation communities.

We initially considered the idea of using a Variational Auto-Encoder (VAE), which is a machine learning model that learns a lower-dimensional representation of the data by reconstructing it from a reduced form that has successfully been used to model complex biological systems [18]. We tried a VAE approach to create our encoding of the data [15]. However, this approach did not produce clear clusters in its latent space due to lack of incentive for this in the loss function. At the time, we were using UMAP: Uniform Manifold Approximation and Projection as a method of visualizing the latent space encoding of the data created by the VAE [8] . We realized that removing the VAE from the pipeline and instead applying the UMAP directly resulted in a dramatic increase in our spatial resolution. This approach produced several clear, distinct clusters (**Fig. 1**).

### 3.2   Dimensionality Reduction of HIV-1 RT Sequences via UMAP and Clustering

1a.
1b.

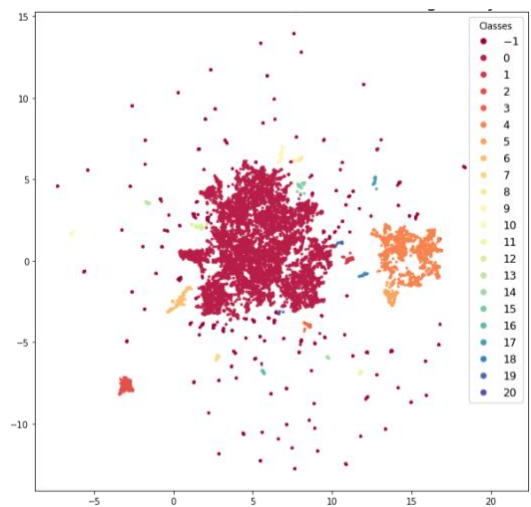
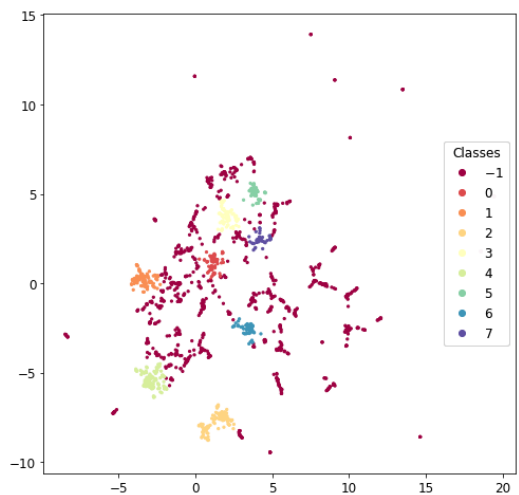

**Figure 1. DBSCAN clustering on Embedding of the training set by UMAP**. This is a low-dimensionality representation of UMAP-mediated dimensionality reduction of our curated HIV-RT sequence database. 1a. Cluster segregation. The sequences in this space are then segregated into 21 clusters using DBSCAN, an unsupervised clustering algorithm based on density. 1b. Second round of UMAP-mediated dimensionality reduction. Cluster 4 is the only cluster in the top figure which provides significant spatial resolution when divided into its two main subclusters.

The UMAP algorithm is a nearest-neighbor graph-based algorithm that performs dimensionality reduction without assumption of linearity [8].   Upon obtaining the UMAP embedding of the sequences, we clustered the sequences in this space using Density-Based Spatial Clustering of Applications with Noise (DBSCAN) [12]. DBSCAN was chosen because it does not require specification of a fixed number of points (sequences) per cluster, instead deriving the optimal number of clusters mathematically. **Figure 1a** details the resulting 20 clusters of sequences obtained and that are further explored in this analysis.

### 3.3 Phylogenetic Relatedness of Clustered Sequences

We performed several tests to assess the biological relevance of the clusters obtained. The first one was to measure the phylogenetic relatedness of the clustered sequences. To this end, we computed a UPGMA tree for each cluster. The maximum depth of the tree represented the distance of the furthest removed descendant of the common ancestor, and thus this was used as a metric for comparison between trees. The results are shown in **Table 1** below. As expected, we observed that the sequences in our clusters tended to be more phylogenetically related than randomly selected sequences, up to 14-fold, although with some exceptions. We also noted that the samples within the clusters also tended to be collected within shorter timeframes than the samples in randomly selected sequences (in selected clusters the timeframe was only 9 years, compared to 30 years for a random sample of similar size). The mean date of collection also shifted substantially depending on the cluster, as late as 2012, compared to 2006 for the random sample. Both the increased phylogenetic relatedness and the increased time frame specificity suggests that the UMAP-derived clusters reflect a biologically-relevant commonality between the sequences.

Table 1: Cluster-by-cluster Phylogenetic Analysis

| Cluster | Size | UPGMA Max Tree Depth | Mutation homogeneity index | Treatment homogeneity index | Years of Collection | Mean year of collection |
|---|---|---|---|---|---|---|
| Random (avg. of 5) | 862 | 0.5030869476 | 0.0162601626 | 0.3571428571 | 1987 to 2017 | 2006 |
| Random (avg. of 5) | 400 | 0.4807792177 | 0.01826484018 | 0.2857142857 | 1987 to 2017 | 2006 |
| Random (avg. of 5) | 50 | 0.3009390475 | 0.02419354839 | 0.5714285714 | 1987 to 2017 | 2006 |
| 0 | 10175 | Too large to compute | 0.009661835749 | 0.3125 | 1987 to 2017 | 2006 |
| 1 | 51 | 0.3282895537 | 0.0412371134 | 0.333 | 1995 to 2017 | 2008 |
| 2 | 375 | 0.5440459029 | 0.07196969697 | 1.000 | 1998 to 2011 | 2006 |
| 3 | 48 | 0.3099992703 | 0.08695652174 | 0.500 | 2002 to 2015 | 2011 |
| 4 | 1737 | 0.3371911882 | 0.0206185567 | 0.200 | 1997 to 2015 | 2010 |
| 5 | 213 | 0.3206148863 | 0.07142857143 | 0.667 | 1995 to 2015 | 2010 |
| 6 | 195 | 0.05779799555 | 0.05940594059 | 0.375 | 1996 to 2016 | 2005 |
| 7 | 51 | 0.02514479978 | 0.15 | n/a | 2001 to 2015 | 2008 |
| 8 | 37 | 0.02633467788 | 0.14 | 1.000 | 2006 to 2015 | 2010 |
| 9 | 62 | 0.03273030533 | 0.1363636364 | 1.000 | 1998 to 2015 | 2008 |
| 10 | 43 | 0.0243315197 | 0.25 | n/a | 2004 to 2015 | 2009 |
| 11 | 48 | 0.0297267636 | 0.1355932203 | n/a | 2004 to 2015 | 2009 |
| 12 | 44 | 0.04062937319 | 0.1230769231 | 1.000 | 1999 to 2015 | 2010 |
| 13 | 41 | 0.0320933801 | 0.1355932203 | n/a | 2002 to 2016 | 2012 |
| 14 | 35 | 0.02136858403 | 0.1219512195 | n/a | 2003 to 2015 | 2010 |
| 15 | 39 | 0.03298188611 | 0.1129032258 | 1.000 | 2002 to 2015 | 2009 |
| 16 | 41 | 0.02498220847 | 0.2173913043 | n/a | 1992 to 2015 | 2010 |
| 17 | 47 | 0.02982838497 | 0.1 | n/a | 2005 to 2015 | 2011 |
| 18 | 31 | 0.04382553236 | 0.07272727273 | 1.000 | 2006 to 2015 | 2011 |
| 19 | 24 | 0.02471175795 | 0.1395348837 | n/a | 2001 to 2015 | 2010 |
| 20 | 15 | 0.3030275172 | 0.1320754717 | 1.000 | 2003 to 2013 | 2011 |

Another way to look at the ability of our analysis to identify related sequences is to look at the representation of all the mutations found in each cluster. To measure the consistency in the representation of mutations within a given cluster, we defined a homogeneity index as $\lambda_{50}/\lambda$ where $\lambda_{50}$ represents the number of mutations with frequency greater than 50% and $\lambda$ represents the total number of mutations observed (see **Methods**). The result is shown in **Table 1**, column 5. As expected, we see a substantial negative correlation between the maximal UPGMA tree length and the homogeneity index (R= -0.57). In other words, the more related the sequences are, the more likely a given mutation is to be highly represented in the cluster.

We noticed that both the phylogenetic relatedness and homogeneity dropped dramatically in seven of the 21 clusters. These included the largest one (cluster 0) and two other medium-sized (clusters 2, 4, and 5). We wondered if an additional round of UMAP analysis for these larger clusters would result in more homogenous subclusters. We only obtained distinct subclusters for cluster 4 (**Fig. 1b**). The UPGMA longest tree and homogeneity indexes for cluster 4 subclusters are listed in **Table 2** and show evidence of phylogenetic relatedness but little mutation homogeneity, suggesting that cluster 4 may represents sequences drifting without clear directionality.

Table 2: Subcluster-by-subcluster Phylogenetic Analysis of Cluster 4

| Cluster | Size | UPGMA Max Tree Depth | Mutation homogeneity index | Treatment homogeneity index | Years of Collection | Mean year of collection |
|---|---|---|---|---|---|---|
| 0 | 76 | 0.03043863312 | 0.07936507937 | 1 | 1998 to 2015 | 2010 |
| 1 | 119 | 0.0355895035 | 0.04938271605 | n/a | 1999 to 2015 | 2009 |
| 2 | 118 | 0.05613080606 | 0.06896551724 | n/a | 2000 to 2015 | 2010 |
| 3 | 88 | 0.04171547765 | 0.08620689655 | n/a | 2000 to 2015 | 2011 |
| 4 | 133 | 0.3253646267 | 0.0625 | 0.5 | 2000 to 2015 | 2011 |
| 5 | 59 | 0.03008323311 | 0.08064516129 | n/a | 2001 to 2015 | 2010 |
| 6 | 63 | 0.02211538217 | 0.08474576271 | n/a | 2001 to 2015 | 2009 |
| 7 | 51 | 0.03636769417 | 0.1090909091 | n/a | 2002 to 2015 | 2010 |

### 3.4 Identifying Drug Selection Correspondence with Clusters

We hypothesized that some of the clusters could identify distinct selections and/or distinct adaptive solutions to selections, as previously shown for networks representing pairwise interactions in TEM beta-lactamase [22, 23]. Since we can assume that most of the viruses included in our database have been under antiretroviral selection since the early nineties, we reasoned that drug treatment may constitute a significant selection driving the evolution of these HIV RT sequences. **Figure 2** shows two side-by-side histograms for each individual cluster: one detailing the frequency of mutations (on the left) and one detailing the drug regimens undergone by the patients from which the viruses were isolated (on the right). Note that for illustration purposes **Figure 2** excludes the mutations which were observed in less than 30% of the sequences.

Interestingly, we observed that eight clusters of sequences identified patients selected by a consistent drug treatment (clusters 2, 5, 8, 9, 12, 15, 18 and 20) (**Figure 2a**). The other 13 clusters are shown in **Figure 2b**.

This is a remarkable result, considering the wide diversity of drug regimens included and that treatment information was not used to generate the clusters.

To quantify the consistency in the treatment, we defined a treatment homogeneity index, similar to the mutation homogeneity index $\mu_{50} / \mu$ where $\mu_{50}$ represents the number of treatments with frequency greater than 50% for a given cluster and $\mu$ represents the total number of treatments observed (see **Methods**). The homogeneity index values are listed in Table 1. All of the above clusters, except cluster 5, produced a homogeneity index of 1. For comparison, a randomly selected control with 50 samples produced a homogeneity index of 0.57.

We also verified that each of these clusters included sequences sourced from various studies, ruling out the possibility of clustering on the basis of individual study design. This adds support to the idea that the algorithm can truly discern the evolutionary effects of selection by particular drug treatments.

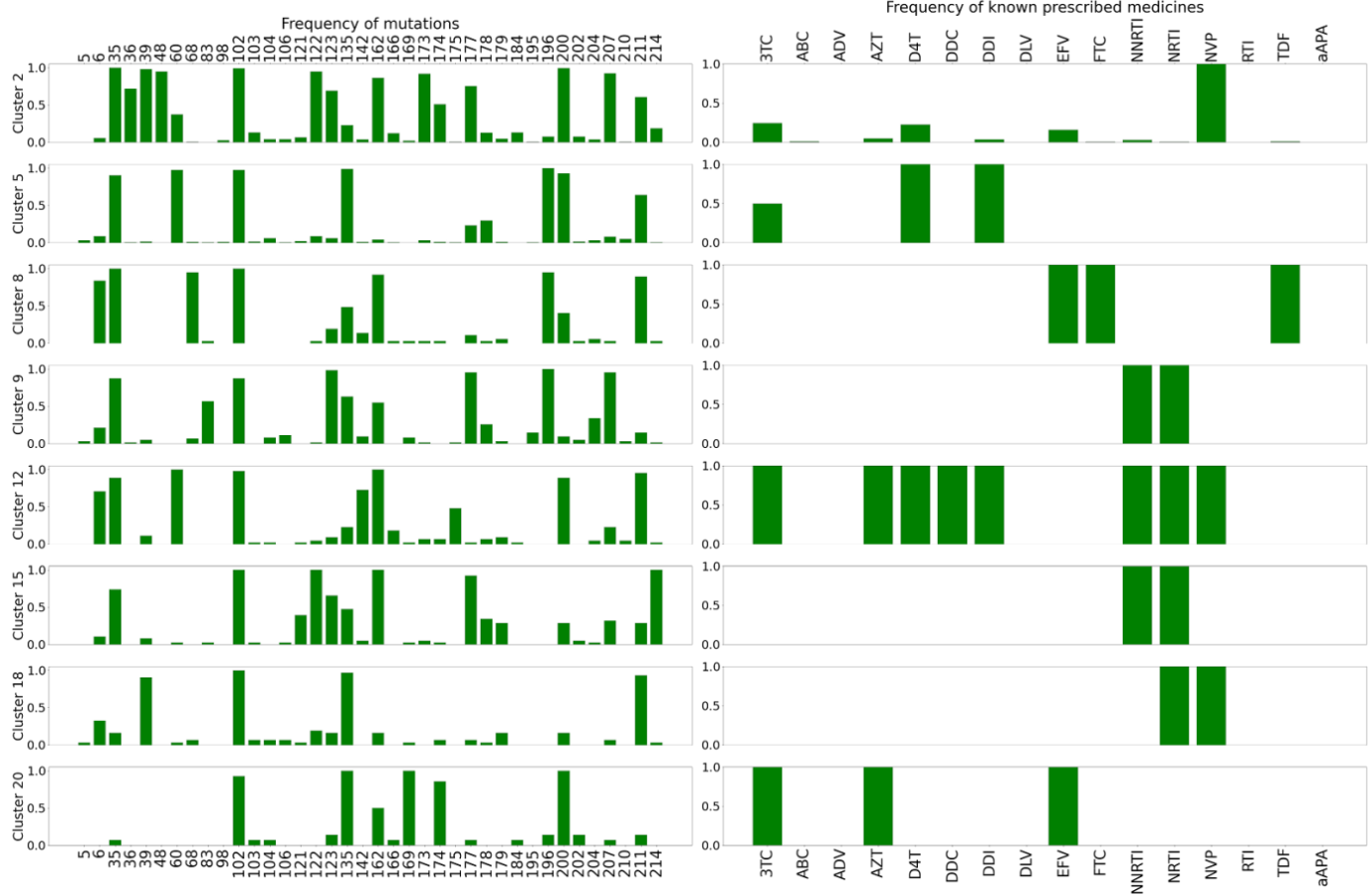

2a.

2b.

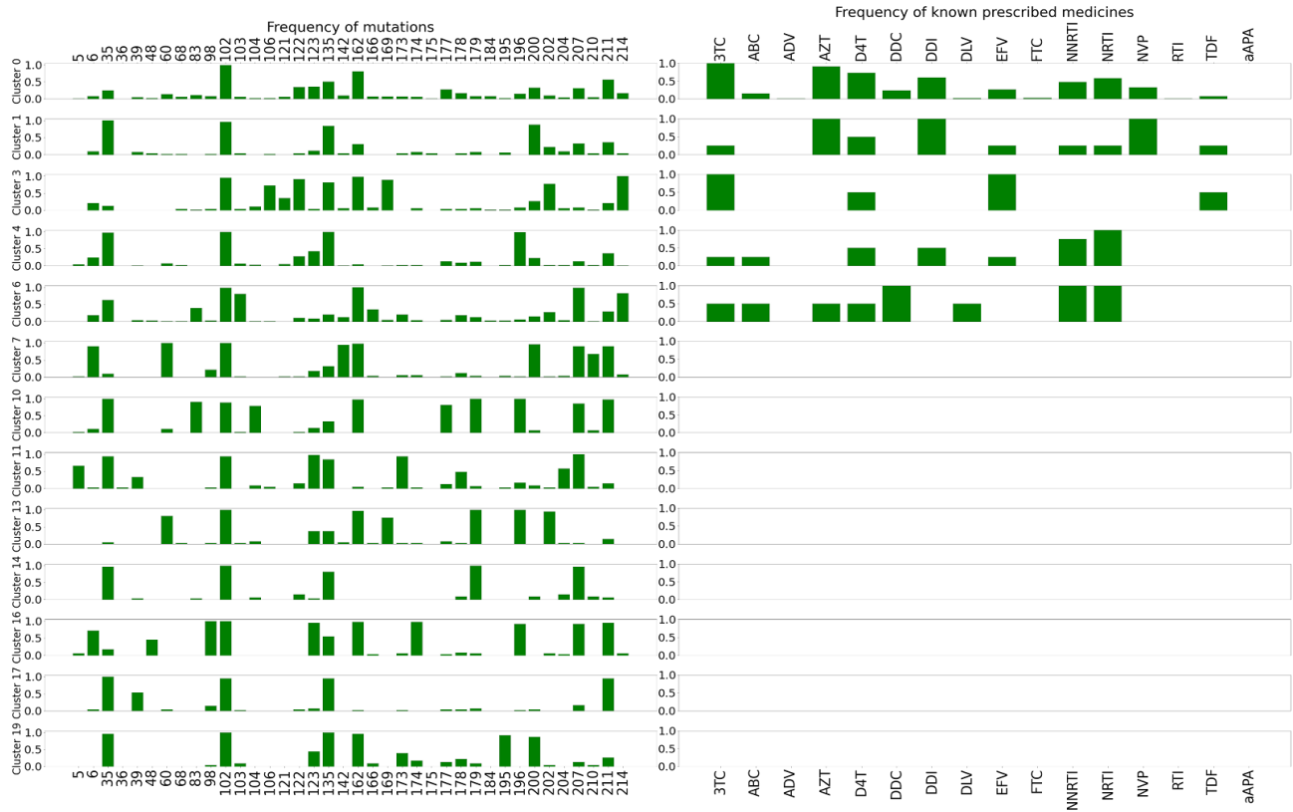

2c.

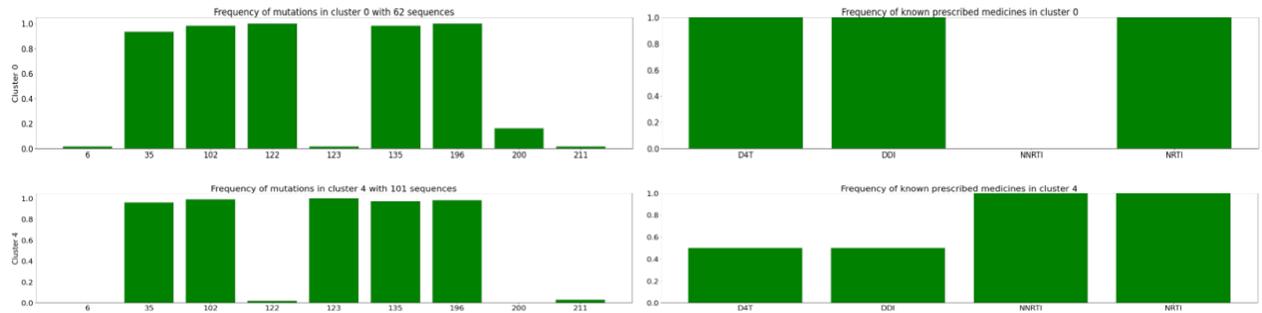

**Figure 2**. **Correspondence between mutation and treatment profiles.** Each row represents a UMAP-derived cluster of HIV-1 RT sequences. Shown are the frequency of mutations by position (left) and the frequency of prescribed drugs in the treatment regimen for each patient (right). **2a**. Clusters with a treatment homogeneity index above a random control. **2b**. Clusters with a treatment homogeneity index below a random control. **2c**. Subclusters 0 and 4 of cluster 4, which show high treatment homogeneity indexes. The correspondence between these clusters and distinct treatment regimens seen in panels **2a** and **2c** suggests that the UMAP clustering can discriminate between different selective pressures effectively.

Strikingly, the mutation homogeneity index correlated strongly with the treatment homogeneity index (R=0.83). We also found a strong treatment concordance for one of the subclusters of cluster 4 (subcluster 0) (**Fig. 2c**). This observation confirms that a repeat UMAP analysis on a heterogeneous cluster can produce subclusters representing concordant treatment.

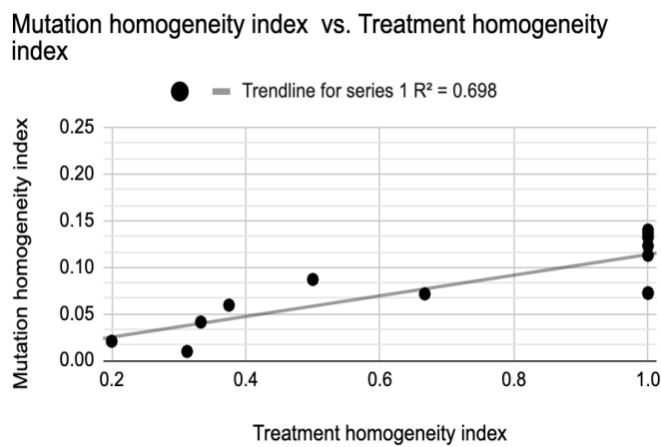

**Figure 3. Correlation between the mutation homogeneity and treatment homogeneity indexes.** The treatment homogeneity index (x-axis) and mutation homogeneity index (y-axis) are plotted. The trendline is shown, and the regression formula is y = 0.111*x + 2.93E-03.

### 3.5   Epistatic Context for Diagnostic Mutations

In a final attempt to demonstrate the link between selected UMAP-derived clusters and drug selection, we looked for a correspondence between mutations that are unique to each cluster and diagnostic mutations for the corresponding selections. To our surprise, we found that the major known drug resistance mutations for NRTI and NNRTI drugs were conspicuously absent as mutations driving cluster classification. Note that mutations present in less than 30% of the sequences for a given cluster were not included in the graph. The main exception was K103X (a NNRTI resistance mutation), which was uniquely found in cluster 6, and L210X (a NRTI resistance mutation, which was uniquely found in cluster 7.

In an effort to show the effect of drug selection, we looked at cluster 2 sequences because we had a good representation of patients sampled before and after treatment (368 out 375 sequences) and because treatment had been consistent across patients (the NNRTI nevirapine, although some of these patients were also treated with NRTIs). For these samples, we called the mutations that were fixed during treatment by identifying mutations that only appeared after treatment. The results of this analysis are shown in **Fig. 4**. In this analysis, we see a high representation of the main NNRTI resistance mutation: K103X and a lower representation of two additional NNRTI mutations: Y181X, G190X. We also see NNRTI mutations represented, notably M184X, D67X, and K70X.

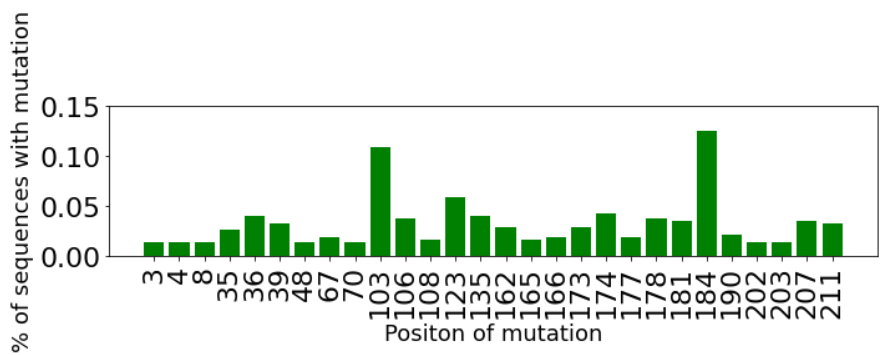

**Fig. 4. Longitudinal analysis of mutations in cluster 2.** Shown are the mutations that only appeared after onset of treatment.

Taken together, these results suggest that for cluster 4, selection is likely driven by drug selection but the network of interactions with the adaptive mutations rather than the adaptive mutation themselves and are more stable and therefore determine the UMAP-derived mutation profile.

## 4 CONCLUSION

Here we used UMAP analysis to embed HIV-RT mutation data in lower-dimensional space and visualized the results using an unsupervised clustering algorithm based on density analysis. We hypothesized that clustering in space would identify evolutionary solutions and/or selections, as previously shown by constructing a network of pairwise interactions for beta-lactamases [22, 23]. We also hypothesized that detecting nonlinear dependencies would increase the resolution of this analysis.

To our knowledge, this is the first time that non-linear dependencies have been included in an analysis of HIV-RT evolution.

Our analysis produced 21 clusters. Here we show three lines of evidence supporting the idea that these clusters represent sequences that are related in a biologically meaningful way. Sequences within a cluster tend to: 1. be phylogenetically related; 2. be sampled closer in time; 3. come from patients undergoing similar treatment. While we have not proven that our clusters represent sequences whose selection is driven by the associated treatment, we have shown a striking correlation between UMAP's ability to identify clusters of homogeneous sequences and the treatment regime of the patients from which these sequences were derived (**Fig. 3**). Note that treatment information was not used as input in this analysis.

To be clear, these are trends and are not found in all the clusters. Repeated UMAP analysis in one of the clusters (cluster 4) showed limited additional resolution, possibly because of drift. Hypermutator viruses have been identified that could accelerate drift. The largest cluster (cluster 0) remains to be resolved possibly because of overlaps between the adaptive solutions needed to respond to complex antiretroviral treatment. Eight clusters have no treatment associated with them, possibly due to an incomplete annotation in our database.

Our analysis produced a mutation profile associated with each cluster. The mutation profiles associated with each cluster may represent the higher-order epistatic context supporting these evolutionary pathways, information that is generally not accessible by other types of mutational co-dependence analyses. Strikingly these profiles contained almost no major DRM. This may partially reflect genetic drift, as common ancestors will be counted as mutations and heavily determine the profile. It may also reflect the fact that these sequences reflect the predominant sequence, and a smaller fraction of sequences with drug resistance mutations may have been missed. This could also mean that the network of interactions with the adaptive mutations rather than the adaptive mutation themselves are more stable and therefore determine the UMAP-derived mutation profile. It will be interesting to investigate these signatures in clusters of mutations known to have evolved within a short period of time in order to derive information about the network of mutant interactions that creates the necessary context for the evolution of specific adaptive solutions.

# Authors' Information

| Your Name | Title* | Research Field | Personal website |
|---|---|---|---|
| Shefali Qamar | Researcher | Computational biology with machine learning applications | https://www.linkedin.com/in/shefali-qamar/ |
| Manel Camps | Full Professor, Doctor | Molecular and computational biology and toxicology | https://www.metx.ucsc.edu/research/lab-sites/camps-laboratory/index.html |
| Jay Kim | Researcher, Doctor | Computational biology | |